\PassOptionsToPackage{table}{xcolor}
\documentclass[sigconf]{acmart}

\usepackage{subcaption}
\usepackage{multirow}
\usepackage{enumitem}
\usepackage{xspace}
\newcommand{\ie}{\emph{i.e.,}\xspace}
\newcommand{\eg}{\emph{e.g.,}\xspace}
\newcommand{\etal}{\emph{et al.}\xspace}

\AtBeginDocument{%
  }

\copyrightyear{2026}
\acmYear{2026}
\setcopyright{cc}
\setcctype{by}
\acmConference[ICMR '26]{International Conference on Multimedia Retrieval}{June 16--19, 2026}{Amsterdam, Netherlands}
\acmBooktitle{International Conference on Multimedia Retrieval (ICMR '26), June 16--19, 2026, Amsterdam, Netherlands}
\acmDOI{10.1145/3805622.3810736}
\acmISBN{979-8-4007-2617-0/2026/06}

\begin{document}

\title{Frozen LVLMs for Micro-Video Recommendation: A Systematic Study of Feature Extraction and Fusion}

\author{Huatuan Sun}
\affiliation{%
  \institution{Nanjing University of Science and Technology}
  \country{China}
}
\email{sunhuatuan@njust.edu.cn}

\author{Yunshan Ma}
\affiliation{%
  \institution{Singapore Management University}
  \country{Singapore}}
\email{ysma@smu.edu.sg}

\author{Changguang Wu}
\affiliation{%
  \institution{Nanjing University of Science and Technology}
  \country{China}
}
\email{changguangwu@njust.edu.cn}

\author{Yanxin Zhang}
\affiliation{%
  \institution{University of Wisconsin-Madison}
  \country{USA}
}
\email{yzhang2879@wisc.edu}

\author{Pengfei Wang}
\affiliation{%
 \institution{GienTech Technology Co., Ltd.}
 \country{China}}
\email{pengfei.wang14@gientech.com}

\author{Xiaoyu Du}
\authornote{Corresponding author}
\affiliation{%
  \institution{Nanjing University of Science and Technology}
  \country{China}}
\email{duxy@njust.edu.cn}


\begin{abstract}

Frozen Large Video Language Models (LVLMs) are increasingly employed in micro-video recommendation (MVR) due to their strong multimodal understanding. However, existing apporches typically deploy LVLMs as fixed black-box feature extractors without systematically comparing alternative representation strategies.
To address this gap, we present the first systematic empirical study on various feature extraction paradigms and integration strategies, along with hierarchical representations from frozen LVLMs for MVR.
Extensive experiments on representative LVLMs reveal that hidden states from multiple decoder layers provide richer and more effective representations for MVR.
Guided by this insight, we propose the Dual Feature Fusion (DFF) Framework, a lightweight approach that adaptively fuses multi-layer representations from frozen LVLMs with ID embeddings. DFF achieves state-of-the-art performance on two real-world micro-video recommendation benchmarks, consistently outperforming strong baselines and providing a principled approach to integrating off-the-shelf large vision-language models into micro-video recommender systems.

\end{abstract}




\maketitle

\section{Introduction}
\label{sec:intro}

The rapid rise of micro-video platforms such as TikTok, Instagram Reels, and YouTube Shorts has made micro-video recommendation a critical component of modern content delivery systems.
It is widely recognized that modeling user preferences from historical interactions, using unique identifiers to represent distinct users and items (IDRec)~\cite{he2017neuralcollaborativefiltering,Kang2018SelfAttentiveSR,hidasi2015gru4rec,yuan2019nextitnet}, has become a mature and industrially successful approach in micro-video recommendation~\cite{gong2022microvideorec1, yu2022microvideorec2, zheng2022microvideorec3}.
Recently, micro-video recommendation has increasingly shifted its focus from purely collaborative signals to hybrid modeling that integrates rich multimodal content semantics~\cite{yuan2023wheretogo,ni2023microlens,zhang2023ninerec}, with approaches leveraging pre-trained large foundational models~\cite{devlin-etal-2019-bert, liu2019roberta, radford2021clip, kim2021viltvisionandlanguagetransformerconvolution} now achieving performance comparable to that of IDRec even in scenarios with abundant user interaction data.
While Large Language Models (LLMs)~\cite{grattafiori2024llama3herdmodels,yang2025qwen3technicalreport,deepseekai2025deepseekv3technicalreport} have significantly advanced diverse tasks that rely on textual modality~\cite{zhang2025notellm2multimodallargerepresentation,zhou2025onerecv2technicalreport, xiang2026mab}, recent works introduce Large Video Language Models (LVLMs) to micro-video recommendation~\cite{ye2025vllmastextencoderandrecommender, De2025vllmcaptionthentextencoder, liu2024recgpt4vmultimodalrecommendationlarge}, leveraging their native multimodal understanding to better capture video semantics and improve recommendation performance. 

Nevertheless, the range of viable strategies for integrating frozen LVLMs into micro-video recommenders has not been systematically evaluated.
First, recent micro-video recommendation methods like MLLM-MSR~\cite{ye2025vllmastextencoderandrecommender} and Nadai~\etal~\cite{De2025vllmcaptionthentextencoder} rely solely
on LVLM-generated captions, implicitly assuming that all recommendation-relevant semantics can be fully captured in text. Yet, this assumption has not been empirically validated; on the contrary, directly leveraging the LVLM's intermediate hidden states, completely bypassing the text generation process, may better preserve recommendation-relevant semantic information.
Second, it is still unclear whether combining LVLM-derived video representations with explicit item identifiers yields better recommendations. Many works ~\cite{ye2025vllmastextencoderandrecommender, cui2022m6recgenerativepretrainedlanguage, hou2023learningvectorquantizeditemrepresentation} tend to directly abandon  these ID features (~\eg replacing item IDs with item contents). However, empirical evidence suggests that LVLMs cannot effectively capture collaborative knowledge without large-scale alignment training ~\cite{zhou2025onerecv2technicalreport, Hu2025alphafuse, huang2024foundationmodelsrecommendersystems} or explicit injection of collaborative signals~\cite{geng2023p5, li2023textneedlearninglanguage, hou2022universalsequencerepresentationlearning}.
Third, most existing methods treat frozen LVLMs as monolithic black-box modules (~\eg caption generators or off-the-shelf feature extractors) without dissecting or leveraging their internal architectural richness. This coarse-grained usage overlooks the potential of selectively fusing multi-granularity representations that may better align with the nuanced demands of micro-video recommendation. Transcending monolithic feature extraction in favor of structured, multi-granularity representation fusion is crucial for harnessing the full potential of LVLMs.
Collectively, these limitations reveal a significant gap in understanding of how frozen LVLMs should be integrated into micro-video recommender systems in a principled and effective manner. This gap motivates three fundamental questions:
(1) What are more effective feature extraction strategies from LVLMs for micro-video recommendation? 
(2) How to effectively integrate these features into recommendation architectures? 
(3) Can we still access richer, task-specific representations by strategically utilizing the inherent structure of frozen LVLMs for recommendation tasks?

To answer the questions, we present a systematic empirical analysis of frozen LVLMs in micro-video recommendation, based on two strong open-source models: MiniCPM-V~2.6/4.5~\cite{yao2024minicpm,yu2025minicpm45} and Video-LaVIT~\cite{jin2024videolavit}.
For clarity and systematic analysis, we modularize the integration of frozen LVLMs into recommender systems into two distinct stages:
feature extraction paradigms $\phi(x_i)$, instantiated by comparing caption-based representations (\ie LVLM-generated text) against hidden-state-based representations (\ie intermediate decoder activations); and integration strategies $\psi(x_i)$, comparing replacement versus fusion with ID embeddings.
Notably, within the feature extraction paradigm, we present a novel layer-wise analysis of intermediate representations across multiple depths of the frozen LVLM, systematically investigating how hierarchical semantic signals at different network layers contribute to micro-video recommendation performance.

By conducting extensive experiments on micro-video recommendation datasets~\cite{ni2023microlens,zhang2023ninerec}, we systematically evaluate design choices across the two distinct stages outlined above. Our empirical analysis reveals three key findings: 
(i) Intermediate hidden states consistently outperform caption-based representations across models and datasets, demonstrating that bypassing text generation preserves richer recommendation-relevant semantics. 
(ii) Fusing LVLM-derived features with item ID embeddings is strictly superior to replacing IDs with content-only representations, confirming the necessity of preserving collaborative signals even when leveraging powerful frozen foundation models~\cite{lin2024recommendersystemsbenefitlarge}.
(iii) The effectiveness of intermediate decoder features varies significantly across layers: middle layers consistently outperform both shallow and deep layers, suggesting an optimal balance between low-level visual details and high-level abstraction. Nevertheless, simple averaging of representations across multiple layers substantially boosts performance, revealing strong complementarity among hierarchical features and highlighting the benefit of multi-granularity fusion.

Building on these findings, we propose a lightweight Dual Feature Fusion (DFF) framework that adaptively integrates multi-scale LVLM features with ID embeddings.
Specifically, DFF extracts hidden states from all decoder layers of the frozen LVLM, projects them into a shared embedding space, and performs a weighted aggregation using learnable global layer weights to emphasize the most informative intermediate layers. 
The resulting content-rich LVLM representation is then adaptively fused with the trainable item ID embedding through a gating mechanism that dynamically balances collaborative signals and multimodal semantics.

In summary, our main contributions are as follows:
\begin{itemize}[leftmargin=0.5cm, itemindent=0cm]
  \item We present the first comprehensive empirical analysis of frozen LVLMs in micro-video recommendation, diagnosing prevalent black-box pitfalls and establishing principled integration guidelines to ensure robust deployment.
  \item We establish three actionable principles: (i) intermediate hidden states consistently outperform caption-based representations; (ii) fusing LVLM features with ID embeddings is strictly superior to replacing them; and (iii) representations from different LVLM layers are complementary, with middle layers being most effective individually, yet multi-layer fusion further boosting performance.
  \item We propose a lightweight Dual Feature Fusion (DFF) framework grounded in these principles, achieving state-of-the-art performance on two representative real-world micro-video recommendation datasets.
\end{itemize}

\section{Related Work}
\begin{figure*}[t!]
    \centering
    \includegraphics[width=0.98\linewidth]{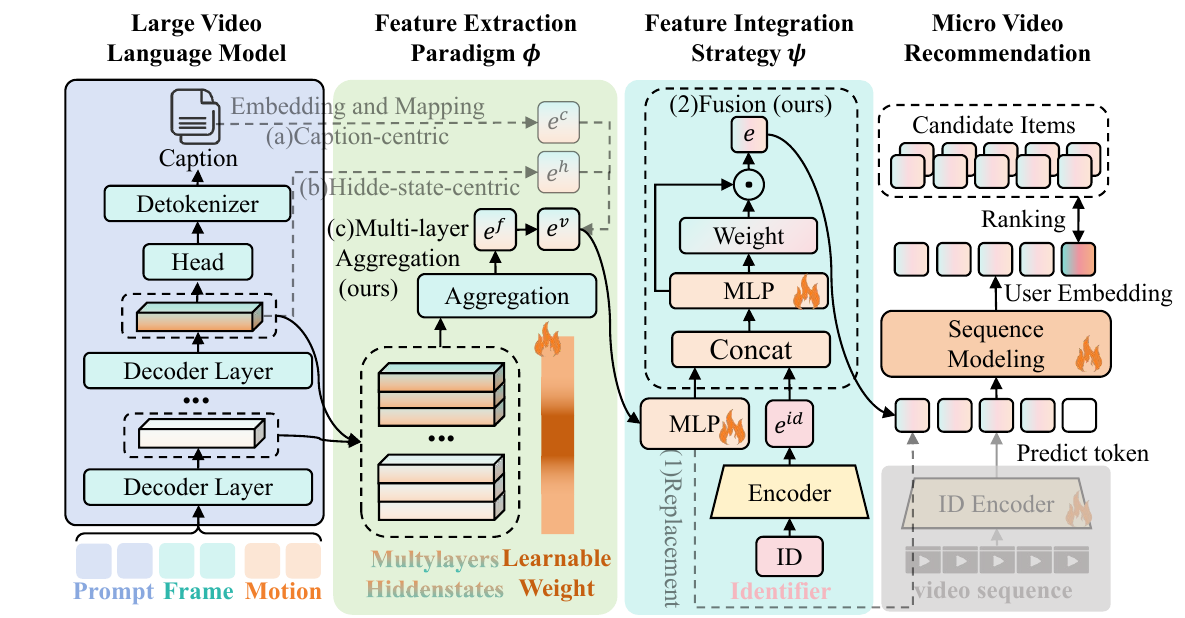}
    \caption{Framework for integrating frozen Large Video Language Models (LVLMs) into micro-video recommendation. The pipeline is characterized by two key design dimensions: \textit{Feature extraction paradigm} $\phi$, including (a) caption-centric, (b) hidden-state-centric, and (c) multi-layer aggregation; and \textit{Integration strategy} $\psi$, which either replaces the ID embedding or adaptively fuses it with LVLM features. 
    Our proposed Dual Feature Fusion (DFF) adopts multi-layer aggregation with adaptive fusion, enabling effective incorporation of both collaborative signals and rich multimodal semantics without LVLM fine-tuning.}
    \label{fig_framework}
\end{figure*}
Our work sits at the intersection of micro-video recommendation, large language models for recommendation, and multi-layer representation fusion. Below, we review related literature along these three axes and highlight the key gaps our study addresses.

\noindent\textbf{Micro-Video Recommendation Systems.}
Early micro-video recommenders primarily relied on collaborative filtering (CF)~\cite{sarwar2001,he2017neuralcollaborativefiltering,Wang2019ngcf,he2020lightgcnsimplifyingpoweringgraph} or sequential modeling~\cite{Kang2018SelfAttentiveSR,hidasi2015gru4rec,yuan2019nextitnet,jiang2020what} based on user–video interaction histories. These methods fall under the ID-based recommendation (IDRec) paradigm, which represents users and items solely via unique identifiers, and have long dominated both academic research and industrial practice in micro-video recommendation.
Nevertheless, the potential of content-aware recommendation remains compelling, particularly for cold-start. Initial attempts to incorporate content leveraged pre-trained CNNs to extract static visual features~\cite{he2015vbprvisualbayesianpersonalized,chen2017acf}, while graph-based models such as MMGCN~\cite{wei2019MMGCN} and MMGCL~\cite{yi2022mmgcl} unified multimodal signals through modality-aware interaction graphs.
Subsequent research in micro-video recommendation has increasingly emphasized fine-grained content understanding, such as modeling user interests at the clip level~\cite{Shang2023FinegrainedUserInterests} or capturing temporal dynamics via time-aware heterogeneous graphs~\cite{Han2022TimeWarping,he2023metapath}.
The release of large-scale benchmarks like MicroLens~\cite{ni2023microlens} and NineRec~\cite{zhang2023ninerec} has further spurred the adoption of advanced pretrained video encoders~\cite{feichtenhofer2019slowfast,tong2022videomaemaskedautoencodersdataefficient}, enabling content-aware methods to achieve performance that exceeds that of ID-based approaches, not only in cold-start settings but also under dense user–video interaction regimes.
However, this performance gain comes at a significant cost, as it requires end-to-end fine-tuning of large video encoders and complex multimodal architectures that are computationally prohibitive for real-world deployment. Moreover, generic representations often introduce spurious correlations that mislead recommendations~\cite{du2022invRL}.
In this work, we aim to develop an effective yet resource-efficient paradigm for micro-video recommendation by systematically exploring how frozen Large Video Language Models (LVLMs) can be best leveraged.

\noindent\textbf{Large Language Models for Recommendation.}
The advent of Large Language Models (LLMs)~\cite{yang2025qwen3technicalreport, deepseekai2025deepseekv3technicalreport, ding2021, sun2019videobertjointmodelvideo} has opened up new possibilities for recommender systems and has been extensively explored~\cite{lin2024recommendersystemsbenefitlarge}.
On one hand, the research community has made significant progress in adapting LLMs to recommender systems~\cite{sheng2025languagerepresentationsrecommendersneed, he2025llm2reclargelanguagemodels, Hu2025alphafuse, dulishuangqing, DMAP}; on the other hand, industry has already deployed several unified-LLM-based recommendation architectures~\cite{zhang2025notellm2multimodallargerepresentation, zhou2025onerecv2technicalreport}.
However, when it comes to micro-video recommendation, the adoption of LVLMs remains largely superficial. 
Most existing works~\cite{De2025vllmcaptionthentextencoder, ye2025vllmastextencoderandrecommender, zheng2025enhancingsequentialrecommenderlarge} merely leverage the textual understanding capabilities of LVLMs in a straightforward manner to enhance item representations for recommendation systems. In contrast, we present a systematic study of frozen LVLM integration in micro-video recommendation, focusing on feature extraction and fusion strategies. We further conduct a fine-grained structural analysis to examine how semantic signals across LVLM layers can be leveraged for recommendation, and provide simple yet effective attempts toward practical use.

\noindent\textbf{Multi-Layer Representation Fusion.}
The practice of fusing representations from multiple layers of deep models is well-established in natural language processing~\cite{devlin-etal-2019-bert, he2021debertadecodingenhancedbertdisentangled, peters-etal-2018-ELMo}, where different layers are shown to encode complementary linguistic features at varying levels of abstraction. Recently, similar layer-wise functional heterogeneity has been observed in multimodal foundation models~\cite{chen2025MultimodalLanguageModelsSeeBetterWhenTheyLookShallower, Lepori2024BeyondtheDoorsofPerceptionVisionTransformersRepresentRelationsBetweenObjects}.  
In contrast, existing recommender systems, particularly those leveraging LVLMs for micro-video recommendation, typically rely on coarse representations (~\eg hidden states from the final decoder layer or generated captions), thereby overlooking the rich, granular semantic hierarchy embedded across intermediate layers. 
To bridge this critical gap, we conduct a comprehensive and systematic analysis of multi-layer representations within LVLMs and explore their adaptive fusion for micro-video recommendation, a strategy that yields significant performance improvements.

\section{Preliminary}
\label{sec:preliminary}
In this section, we formalize the micro-video sequential recommendation task under a resource-efficient paradigm: the Large Video Language Model (LVLM) is kept frozen as an off-the-shelf feature extractor, and all trainable parameters are confined to the recommendation architecture. This setup enables strong content understanding without costly LVLM fine-tuning. 

\noindent\textbf{Task Formulation.}
We investigate frozen Large Video Language Models (LVLMs) under the standard micro-video sequential recommendation setting. Let $\mathcal{U}$ and $\mathcal{V}$ denote the sets of users and micro-videos, respectively. The interaction history of a user $u \in \mathcal{U}$ is an ordered sequence $\mathcal{S}_u = [v_1, v_2, \dots, v_{t}]$, where $v_i \in \mathcal{V}$ is the $i$-th interacted video. The goal is to predict the user's next preferred video $v_{t+1}$ given $\mathcal{S}_u$.
A recommender model $R(\cdot)$ learns a scoring function over the candidate set $\mathcal{V}$. In content-aware systems, each video $v_i$ is associated with rich raw input data (~\eg its video file $\mathbf{x}$). An embedding function maps this raw content to a dense vector representation $\mathbf{e}$. Consequently, the prediction is made based on the sequence of these embeddings:
\begin{equation}
    p(v_{t+1} \mid \mathcal{S}_u) = R([\mathbf{e}_1, \dots, \mathbf{e}_{t}]).
    \label{eq:rec_task}
\end{equation}

\noindent\textbf{Integrating Frozen LVLMs: A Decomposed Framework.}
To leverage the powerful multimodal understanding of LVLMs without fine-tuning, we decompose the video embedding function into two distinct modules, as illustrated in Figure~\ref{fig_framework}:
\begin{itemize}[leftmargin=*,nosep]
    \item \textbf{Frozen Feature Extractor $\phi(\cdot)$}: With a frozen pre-trained LVLM, it takes the raw video input $\mathbf{x}_i$ and outputs a content-rich feature vector$\mathbf{e}^v$. The choice of how to extract this feature from the LVLM constitutes our first key design dimension.
    
    \item \textbf{Trainable Integrator $\psi(\cdot)$}: This lightweight, learnable module combines the frozen LVLM feature $\mathbf{e}^v$ with the ID embedding $\mathbf{e}^{\text{id}}$. The choice of how to integrate these two representations constitutes our second key design dimension.
\end{itemize}
The final video embedding $\mathbf{e}_i$ is obtained by composing these two modules:
\begin{equation}
    \mathbf{e}_i = \psi\big( \phi(\mathbf{x}_i),\, \mathbf{e}_i^{id} \big).
    \label{eq:embedding}
\end{equation}

Under this decomposed framework, the entire problem of LVLM integration reduces to making principled choices for $\phi_{\text{LVLM}}$ and $\psi_{\theta}$. 

\noindent\textbf{Preliminaries on LVLMs.}
\noindent\textbf{Data Preprocessing.}
We use the original MP4 video files from the datasets as input. 
For each video, frames are extracted and preprocessed using the official pipeline provided by the respective LVLM. 
Consistent with the design constraints of current LVLMs, we uniformly sample up to 16 frames per video, which is the maximum input length supported by all evaluated models.

\noindent\textbf{LVLM Selection.}
We evaluate two representative frozen Large Video Language Models (LVLMs), each exhibiting distinct approaches to video understanding:
\begin{itemize}[left=0pt]
    \item \textbf{MiniCPM-V 2.6}~\cite{yao2024minicpm}: treats videos as sequences of independent static images and lacks explicit temporal modeling;
    \item \textbf{MiniCPM-V 4.5}~\cite{yu2025minicpm45}: integrates multi-frame visual inputs via a 3D Resampler module to produce a unified spatio-temporal representation aligned with text;
    \item \textbf{Video-LaVIT}~\cite{jin2024videolavit}: extracts keyframes and motion cues, then compresses the video into compact tokens through dedicated Tokenizer and Detokenizer modules, enabling fine-grained joint video–language modeling in a cross-modal Transformer.
\end{itemize}

\noindent\textbf{Prompt Design.}
During inference, all LVLMs are prompted with the following task-specific instruction:
\begin{quote}
``You are watching \textbf{a micro-video on a social media platform}. Please analyze the video content for application in a \textbf{micro-video recommendation system}. Write one coherent paragraph describing the scene, people or objects, main actions, and the style or type.''
\end{quote}

\section{Systematic Exploration of LVLM in Micro-Video Recommendation}
\label{sec:exploration}

To establish principled guidelines for leveraging frozen LVLMs in micro-video recommendation, we conduct a systematic empirical study along the two design dimensions introduced in Section~\ref{sec:preliminary}. 

\begin{table}[t]
    \centering
    \caption{Performance comparison of different LVLM models and feature integration strategies on the MicroLens dataset.}
    \setlength{\tabcolsep}{8pt}
    \label{tab_exploring}
    \begin{tabular}{ccccc}
        \toprule
        \multirow{2}{*}[-2pt]{Model} & \multicolumn{4}{c}{MicroLens}\\
        \cmidrule(r){2-5}
        & H@10 & N@10 & H@20 & N@20\\
        \midrule
        \rowcolor{gray!10} \multicolumn{5}{c}{Baseline} \\
        ID & 0.0909 & 0.0517 & 0.1278 & 0.0610\\
        VIDRec & 0.0799 & 0.0415 & 0.1217 & 0.0520\\
        VideoRec & 0.0948 & 0.0515 & 0.1364 & 0.0619\\
        \rowcolor{gray!10} \multicolumn{5}{c}{VLLM Feature Usage Strategies} \\
        Replacement & 0.0665 & 0.0375 & 0.1123 & 0.0487\\
        Fusion  & 0.0975 & 0.0545 & 0.1353 & 0.0640\\
        \bottomrule
    \end{tabular}
\end{table}

\subsection{Feature Integration Strategy $\psi(\cdot)$}
For Integration Strategy, we investigate how to combine the LVLM feature $e^{v}$ with the collaborative signal from the ID embedding $e^{\text{id}}$:

\noindent\textbf{Replacement:} The LVLM feature $e^{v}$ (projected into the behavioral embedding space) completely replaces the conventional ID embedding $e^{\text{id}}$ in the recommender backbone. This strategy assumes that the LVLM's rich content semantics subsume the need for explicit collaborative signals.

\noindent\textbf{Fusion:} The LVLM feature $e^{v}$ is combined with the ID embedding $e^{\text{id}}$ through a lightweight, trainable gating mechanism. Formally, the final representation $e$ is computed as:
\begin{align}
    g &= \sigma\!\left( \mathrm{MLP}\big([e^{\text{id}}; e^{v}]\big) \right), \label{eq:gate} \\
    e &= g \cdot e^{\text{id}} + (1 - g) \cdot e^{v}, \label{eq:fused_emb}
\end{align}
where $\sigma(\cdot)$ denotes the sigmoid function. This adaptive fusion allows the model to dynamically balance collaborative (ID-based) and semantic (LVLM-based) information for each item.

Specifically, we adopt Video-LaVIT~\cite{jin2024videolavit} as the frozen LVLM and use the hidden states from its last decoder layer as the LVLM-derived video representation. In addition, we include three representative baselines from MicroLens~\cite{ni2023microlens} for comparison: \textbf{ID}: represents each video solely by its learnable ID embedding, capturing only collaborative signals from user-item interaction sequences while ignoring visual content; \textbf{VIDRec}: encodes raw video frames using a frozen video encoder and concatenates the resulting visual features with the video ID embedding; \textbf{VideoRec}: fine-tunes the video encoder on the recommendation task and then fuses its output with the ID embedding via concatenation.
These methods provide a spectrum of approaches, ranging from pure collaborative filtering to content-aware modeling, against which we evaluate our LVLM-based integration strategies.

Table~\ref{tab_exploring} demonstrates that LVLM features do not replace ID embeddings but substantially enhance them when properly fused. Specifically, the Replacement strategy performs even worse than the ID-only baseline, which relies solely on collaborative signals without any visual content, highlighting the irreplaceable value of ID embeddings in capturing user–item interaction patterns.
In contrast, the \textit{Fusion} approach significantly outperforms VIDRec, demonstrating that LVLM-derived features provide rich and complementary semantic information beyond what conventional frozen video encoders can offer. Remarkably, despite using only a frozen LVLM and a simple concatenation-based integration, Fusion also surpasses VideoRec, the method that fine-tunes the video encoder on the recommendation task. 
This result underscores the strong potential of off-the-shelf LVLMs in recommendation systems: when properly integrated, they enable superior performance without costly task-specific training.

\begin{figure}[htbp]
    \includegraphics[width=1.0\linewidth]{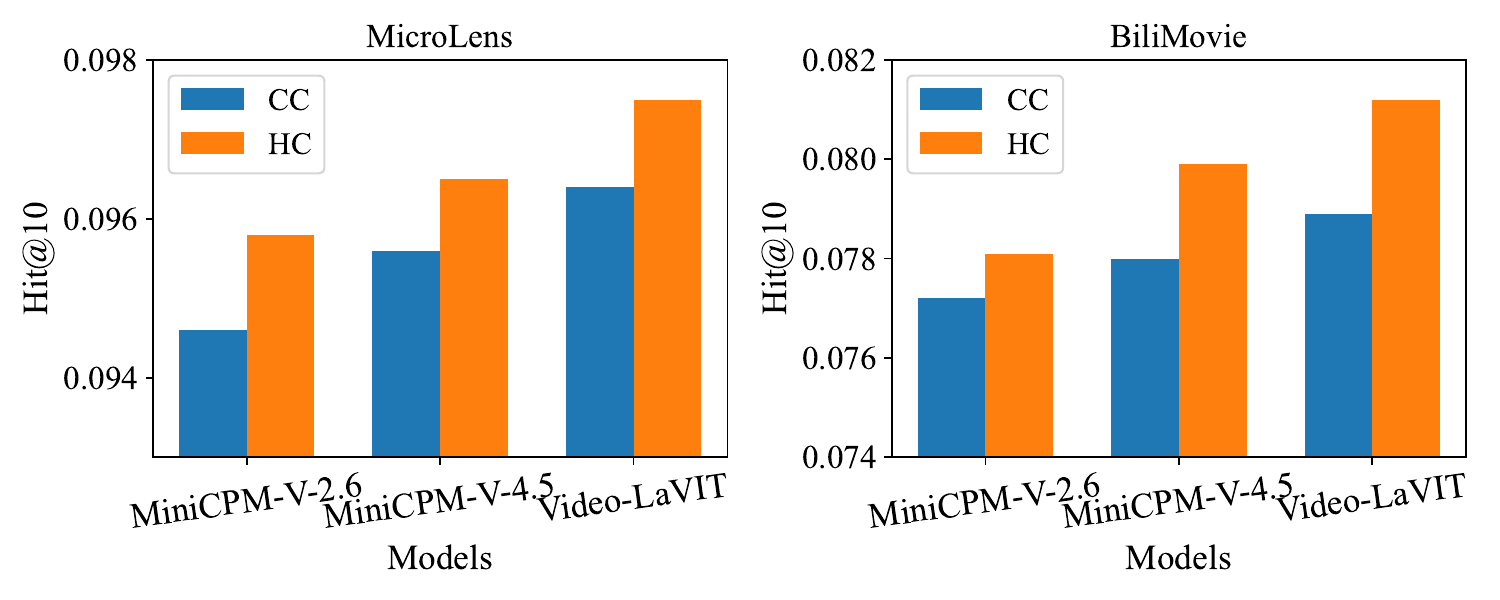}
    \caption{Performance comparison of CC (Caption-Centric) and HC (Hidden-State-Centric) feature extraction strategies, that adopts the generated captions and the hidden states of LVLM decoders as item representations, respectively.}
    \label{fig_empirical_experiments_feature_extract}
\end{figure}
\subsection{Feature Extraction Paradigm $\phi(\cdot)$}
For feature extraction paradigm ($\phi$), We compare two distinct methods to derive a semantic feature vector from the frozen LVLM given a raw video $x_i$:

\noindent\textbf{Caption-Centric (CC):} Following prior work~\cite{De2025vllmcaptionthentextencoder,ye2025vllmastextencoderandrecommender}, the LVLM is used as a sophisticated caption generator. Given a prompt, it produces a descriptive textual narrative of the video content. This generated caption is then encoded into a fixed-dimensional vector $e^{c}$ using a state-of-the-art text encoder (\eg the OpenAI's text-embedding-3-large). This approach treats language as an intermediary proxy for visual semantics.

\noindent\textbf{Hidden-State-Centric (HC):} We bypass text generation entirely and directly extract the intermediate hidden states from the LVLM's decoder. Specifically, we use the hidden state from the final decoder layer as the content-aware representation $e^{h}$. This method leverages the LVLM's internal, dense semantic representation before it is projected into discrete tokens.

We evaluate both paradigms using two representative frozen LVLMs, MiniCPM-V and Video-LaVIT, on the MicroLens~\cite{ni2023microlens} and Bili\_Movie~\cite{zhang2023ninerec} datasets.
As shown in Figure~\ref{fig_empirical_experiments_feature_extract}, the hidden-state-centric features outperform the caption-centric features, and the choice of LVLM model also leads to considerable differences in performance. From these observations, we draw the following conclusions: (1) employing LVLM as a video encoder shows great potential for micro-video recommendation; (2) directly using hidden states as representations of video content is more effective than utilizing captions generated by the LVLM; and (3) the capability of LVLM to process video is crucial for providing better features to micro-video recommendation systems.

\begin{figure}[htbp]
    \centering
    \includegraphics[width=0.95\linewidth]{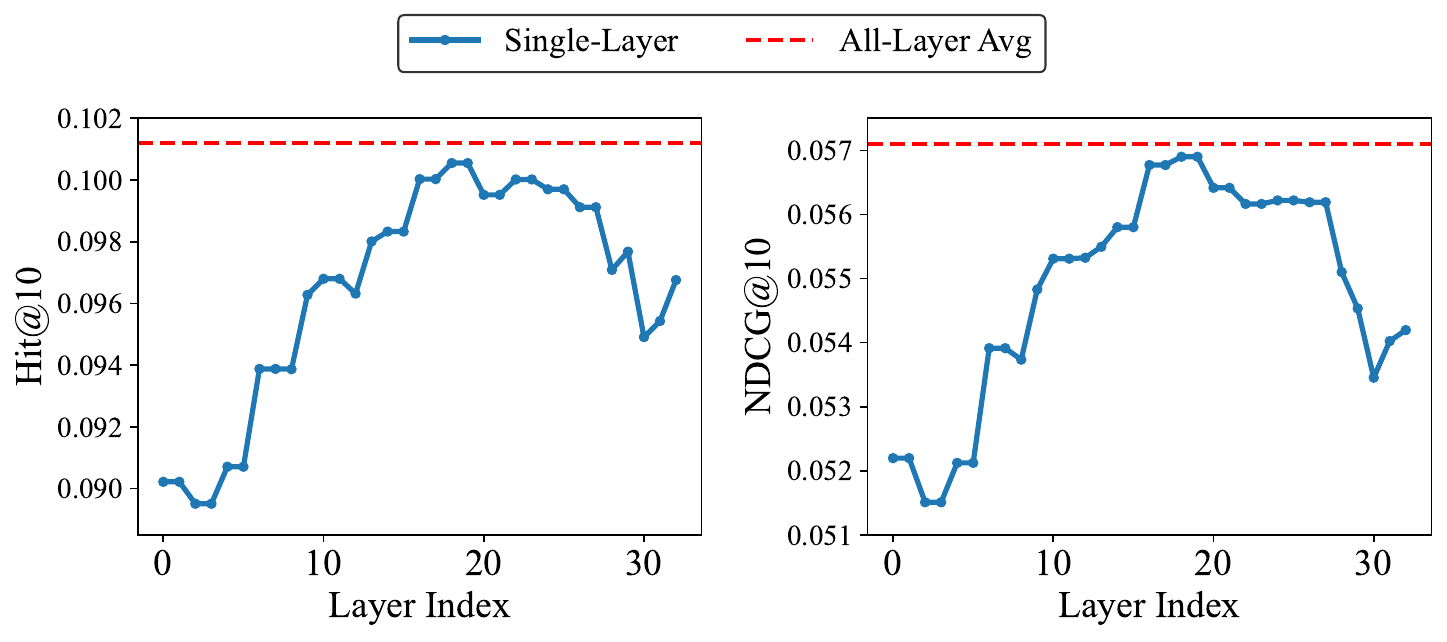}
    \caption{Performance of layer-wise features from Video-LaVIT on MicroLens. The red dashed line shows the result of uniformly averaging hidden states across all decoder layers (Hit@10: 0.1012, NDCG@10: 0.0571).}
    \label{fig_empirical_experiments_different_layers}
\end{figure}

\subsection{Features in Different Decoder Layers}
\label{sec:layer_analysis}
While most existing approaches extract features from a single layer, typically the final layer, of frozen large language models, recent studies~\cite{chen2025MultimodalLanguageModelsSeeBetterWhenTheyLookShallower, Lepori2024BeyondtheDoorsofPerceptionVisionTransformersRepresentRelationsBetweenObjects} have shown that representations from different layers encode complementary semantic signals at varying levels of abstraction.
Motivated by these critical observations, we perform a comprehensive and systematic layer-wise analysis on the sophisticated decoder architecture of \textbf{Video-LaVIT}, aiming to meticulously examine how its nuanced video understanding capability progressively evolves and transforms across varying depths within the context of micro-video recommendation.

We conduct this analysis on the MicroLens dataset~\cite{ni2023microlens}. Specifically, each micro-video is fed into Video-LaVIT, and the hidden states $\{h^{(l)}\}_{l=1}^L$ from all $L$ decoder layers are extracted. These layer-specific representations $h^{(l)}$ are first projected into the behavioral embedding space and then serve as the content-aware embedding $e^{v}$ in the gating-based fusion mechanism (Section~\ref{sec:preliminary}), where they are adaptively combined with the corresponding item ID embedding $e^{\text{id}}$. We evaluate their effectiveness using two standard ranking metrics: \textbf{Hit@10} and \textbf{NDCG@10}. This yields a performance curve that reveals how recommendation quality varies with layer depth. 
In addition, we investigate a lightweight yet effective fusion strategy: uniformly averaging the hidden states from multiple selected layers before projecting and fusing them with ID embeddings.

As shown in Figure~\ref{fig_empirical_experiments_different_layers}, the recommendation performance varies significantly across decoder layers. 
In both Hit@10 and NDCG@10, we observe a consistent trend: shallow layers (\eg layers 1–10) yield suboptimal representations due to insufficient semantic abstraction, while the deepest layers (\eg layers 30–33) exhibit slight performance degradation, possibly caused by over-smoothing or task-irrelevant linguistic priors. 
Notably, the middle layers (approximately layers 16–22) consistently achieve peak performance, suggesting that they strike an optimal balance between low-level visual grounding and high-level semantic reasoning.
Furthermore, simple uniform averaging of hidden states from all decoder layers, without any learnable fusion weights, yields a Hit@10 of 0.1012 and NDCG@10 of 0.0571, which surpasses the best single-layer results (0.1005 and 0.0569, respectively).
This highlights the high degree of complementarity inherent in representations from different depths, demonstrating that even a lightweight fusion strategy can effectively harness this diversity to significantly enhance the quality of video representation for micro-video recommendation.

\section{Our Approach: DFF}
\label{sec:dual_fusion}

Motivated by our empirical findings, namely that (i) intermediate Large Video Language Model (LVLM) representations preserve richer semantics than distilled captions, (ii) ID embeddings encode irreplaceable collaborative signals, and (iii) multi-layer decoder states are complementary, we propose a lightweight yet effective Dual Feature Fusion (\textbf{DFF}) framework for integrating frozen LVLMs into micro-video recommenders.
DFF unifies two key design principles into a trainable pipeline: adaptive aggregation of multi-granularity LVLM features across decoder layers and fusion of these content-rich representations with collaborative ID embeddings (~\ie Multi-layer Aggregation and Fusion in Figure~\ref{fig_framework}).

\noindent\textbf{Multi-Layer Aggregation.}
Instead of relying on a single decoder layer or uniform averaging (as explored in Section~\ref{sec:layer_analysis}), we introduce learnable layer weights to dynamically emphasize the most informative layers for recommendation. 
Given a micro-video $x$, we extract hidden states $\{h^{(l)}\}_{l=1}^L$ from all $L$ decoder layers of a frozen LVLM (\eg Video-LaVIT). 
We compute a weighted combination:
\begin{align}
    e^v = \sum_{l=1}^L \alpha_l \cdot h^{(l)},
\end{align}
where $\boldsymbol{\alpha} = [\alpha_1, \dots, \alpha_L]^\top$ are \textit{global} learnable weights shared across the entire video corpus and collectively optimized end-to-end alongside the recommender.
This design balances model capacity and generalization by avoiding per-item overfitting while capturing cross-layer complementarity.

\noindent\textbf{Adaptive Fusion with Collaborative ID Embeddings.}
Following the finding that the ID embedding is irreplaceable (Section~\ref{sec:exploration}), we combine the aggregated LVLM feature $e^v$ with collaborative signals. Specifically, we adopt the same adaptive fusion mechanism: given the ID embedding $e^{\text{id}} = \mathrm{IDEncoder}(i)$ and the aggregated LVLM feature $e^v$, the final representation $e$ is computed as in Equations~\eqref{eq:gate}-\eqref{eq:fused_emb}.
The resulting fused embedding $e$ is then fed into a sequential recommender backbone (~\eg SASRec~\cite{Kang2018SelfAttentiveSR}) to model user behavior and predict the next micro-video. 
Notably, the entire pipeline, including the layer weights $\boldsymbol{\alpha}$, projection head, and gating MLP, is jointly optimized via the recommendation loss (~\ie cross-entropy loss).
Despite its simplicity, DFF effectively leverages the full potential of frozen LVLMs without fine-tuning, making it highly practical for real-world deployment.

\section{Experiments and Analysis}
In this section, we quantitatively evaluate the proposed Dual Feature Fusion (DFF) framework on two micro-video recommendation benchmarks and provide in-depth analyses of key practical factors.

\subsection{Datasets and Evaluation Metrics}
In this work, we utilize two primary datasets: MicroLens~\cite{ni2023microlens} and Bili\_Movie (the Bilibili movie domain of the NineRec benchmark\cite{zhang2023ninerec}). The MicroLens dataset serves as a large-scale benchmark for micro-video recommendation, comprising millions of user–video interaction records along with rich multimodal information, including videos, textual descriptions, audio tracks, and cover images. MicroLens has been widely adopted in both micro-video recommendation tasks and multimodal learning research, providing a standard and comprehensive testbed for algorithm evaluation and model development. We mainly utilize the raw video data from the MicroLens dataset as model input in this work. We follow the dataset’s official splitting protocol, adopting a leave-one-out strategy, where the last item in each user’s interaction sequence is used for evaluation, the second-to-last for validation, and the remaining items for training. Hit Rate (HR@N)~\cite{he2017neuralcollaborativefiltering} and Normalized Discounted Cumulative Gain (NDCG@N)~\cite{j2002ndcg} are used as our ranking evaluation metrics. 
The original Bili\_Movie dataset also offers multimodal data and user–video interaction records for micro-videos. We further processed this dataset by downloading and preprocessing the original video content using the provided URLs. 
Notably, to ensure a consistent and comparable experimental setup, we adopt the same preprocessing pipeline, data splitting strategy, and evaluation protocol on Bili\_Movie as established in the MicroLens benchmark~\cite{ni2023microlens}.

\subsection{Implementation Details}
All large video language models (LVLMs) inference is performed on a single NVIDIA A800 GPU, while training of the recommendation models is conducted on an NVIDIA GeForce RTX 4090 GPU unless otherwise specified. 
We optimize the model using the AdamW optimizer~\cite{loshchilov2019adamw} with a fixed weight decay of 0.1 and a batch size of 512. 
The learning rate and intermediate embedding dimension are treated as hyperparameters and selected from $\{1 \times 10^{-4}, 1 \times 10^{-5}, 1 \times 10^{-6}\}$ and $\{512, 1024, 2048\}$, respectively, based on validation performance. 
Training employs early stopping with a patience of 5 epochs, up to a maximum of 100 epochs.
\begin{table*}[t]
    \centering
    \normalsize    
    \caption{Performance comparison of models on MicroLens and Bili\_movie datasets. Bold denotes the best result for each column, and underlined values indicate the second-best result.
    An asterisk\textsuperscript{*} indicates that the result is statistically significant ($p < 0.05$) according to paired t-tests conducted over five independent runs with different random seeds.}
    \setlength{\tabcolsep}{8pt}
    \begin{tabular}{ccccccccc}
        \toprule
        \multirow{2}{*}[-2pt]{Model} & \multicolumn{4}{c}{MicroLens} & \multicolumn{4}{c}{Bili\_movie}\\
        \cmidrule(r){2-5}\cmidrule(l){6-9}
        & H@10 & N@10 & H@20 & N@20 & H@10 & N@10 & H@20 & N@20\\
        \midrule
        \rowcolor{gray!10} \multicolumn{9}{c}{IDRec (official baselines from MicroLens)} \\
        SASRec & 0.0909 & 0.0517 & 0.1278 & 0.0610 & 0.0760 & 0.0441 & 0.1074 & 0.0520\\
        GRU4Rec & 0.0782 & 0.0423 & 0.1147 & 0.0515 & 0.0621 & 0.0427 & 0.1012 & 0.0462\\
        NextItNet & 0.0805 & 0.0442 & 0.1175 & 0.0535 & 0.0691 & 0.0433 & 0.1056 & 0.0498\\
        \rowcolor{gray!10} \multicolumn{9}{c}{VideoRec (official baselines from MicroLens)} \\
        SASRec & 0.0948 & 0.0515 & 0.1364 & 0.0619 & 0.0769 & 0.0449 & 0.1102 & 0.0523\\
        GRU4Rec & 0.0954 & 0.0517 & 0.1377 & 0.0623 & 0.0770 & 0.0451 & 0.1132 & 0.0527\\
        NextItNet & 0.0862 & 0.0466 & 0.1246 & 0.0562 & 0.0712 & 0.436 & 0.1071 & 0.0506\\
        \rowcolor{gray!10} \multicolumn{9}{c}{Caption-centric Feature Extraction and ID Embedding Fusion} \\
        MiniCPM-V 2.6 & 0.0946 & 0.0533 & 0.1317 & 0.0627 & 0.0772 & 0.0449 & 0.1109 & 0.0513\\
        MiniCPM-V 4.5 & 0.0956 & 0.0537 & 0.1333 & 0.0632 & 0.0780 & 0.0453 & 0.1130 & 0.0524\\
        Video-LaVIT & 0.0964 & 0.0541 & 0.1334 & 0.0634 & 0.0789 & 0.0450 & 0.1144 & 0.0533\\
        \rowcolor{gray!10} \multicolumn{9}{c}{Hidden-state-centric Feature Extraction and ID Embedding Fusion} \\
        MiniCPM-V 2.6 & 0.0958 & 0.0540 & 0.1341 & 0.0636 & 0.0781 & 0.0452 & 0.1139 & 0.0522\\
        MiniCPM-V 4.5 & 0.0965 & 0.0542 & 0.1343 & 0.0637 & 0.0799 & 0.0449 & 0.1142 & 0.0531\\
        Video-LaVIT & \underline{0.0975} & \underline{0.0545} & \underline{0.1353} & \underline{0.0640} & \underline{0.0812} & \underline{0.0454} & \underline{0.1151} & \underline{0.0539}\\
        \rowcolor{gray!10} \multicolumn{9}{c}{Our Method} \\
        DFF \textit{(ours)} & \textbf{  0.1020$^{*}$} & \textbf{  0.0577$^{*}$} & \textbf{  0.1414$^{*}$} & \textbf{  0.0676$^{*}$} & \textbf{  0.0839$^{*}$} & \textbf{  0.0471$^{*}$} & \textbf{  0.1156$^{*}$} & \textbf{  0.0551$^{*}$}\\
        \it Imp. & 4.61\% & 5.87\% & 4.50\% & 5.62\% & 3.32\% & 3.74\% & 0.43\% & 2.22\%\\
        \bottomrule
    \end{tabular}
    \label{tab_experiments}
\end{table*}
\subsection{Overall Performance}
Following the empirical analysis in Section~\ref{sec:exploration}, we now evaluate the complete Dual Feature Fusion (DFF) framework using multiple LVLMs (~\ie MiniCPM-V 2.6/4.5 and Video-LaVIT). 
As Table~\ref{tab_experiments} shows, DFF achieves the best performance with H@10 of 0.1020 and N@10 of 0.0577 on MicroLens, and H@10 of 0.0839 and N@10 of 0.0471 on Bili\_Movie. Compared to the strongest official baseline from MicroLens (GRU4Rec with VideoRec, H@10: 0.0954), DFF achieves a 6.91\% relative gain, validating that a carefully designed fusion of multi-layer LVLM features and ID embeddings can outperform end-to-end trained video recommendation models even while keeping the LVLM entirely frozen. Moreover, DFF also outperforms Video-LaVIT with hidden-state-centric feature extraction and ID embedding fusion (~\ie the best frozen-LVLM variant in Section~\ref{sec:exploration}, H@10: 0.0975), by 4.61\%, demonstrating the effectiveness of multi-layer feature aggregation over single-layer representations. Similar gains are observed on Bili\_Movie, where DFF surpasses GRU4Rec with VideoRec (H@10: 0.0770) by 8.96\% and the best hidden-state-centric frozen variant (H@10: 0.0812) by 3.32\%. This consistent superiority across datasets underscores that adaptive multi-layer fusion is a reliable and transferable strategy for leveraging frozen vision-language models in recommendation.

\subsection{In-depth Analysis}
\begin{table}[t]
    \centering
    \caption{Impact of prompt engineering on VLLM-based micro-video recommendation performance.}
    \begin{tabular}{ccccc}
        \toprule
        Method & H@10 & N@10 & H@20 & N@20\\
        \midrule
        w/o video rec prompt & 0.0953 & 0.0533 & 0.1331 & 0.0628\\
        w/ video rec prompt & 0.0975 & 0.0545 & 0.1353 & 0.0640\\
        \bottomrule
    \end{tabular}
    \label{tab_prompt}
\end{table}
Beyond feature extractor and integrator choices, the performance is also sensitive to practical deployment factors such as prompt formulation and input modality, which are often overlooked in recommendation literature. 
To offer an in-depth analysis of real-world applicability, we systematically investigate these factors.
\subsubsection{The Impact of Prompt Engineering.}
We design a prompt for micro-video recommendation by incorporating task-specific keywords, which is referred to as the \textbf{video rec prompt}. 
As shown in Table~\ref{tab_prompt}, we compare two settings: \textbf{w/o video rec prompt}, where we only indicate that the input is a video in order to obtain the video content representation; and \textbf{w/ video rec prompt}, where we prepend a prompt specifically designed for the recommendation scenario before inputting the video to guide the LVLM towards extracting features suitable for recommendation. The two prompts are as follows:

\begin{itemize}[leftmargin=0.5cm, itemindent=0cm]
\item \textbf{w/o video rec prompt:} “You are watching a video. Write one coherent paragraph describing the scene, people or objects, main actions, and the style or type.”
\item \textbf{w/ video rec prompt:} “You are watching \textbf{a micro-video on a social media platform}. Please analyze the video content for application in a \textbf{micro-video recommendation system}. Write one coherent paragraph describing the scene, people or objects, main actions, and the style or type.”
\end{itemize}

Experimental results show that explicit recommendation-oriented prompts consistently improve performance across all metrics. This confirms that task-specific prompts effectively align LVLM representations with the discriminative nature of recommendation tasks, highlighting the critical role of prompt engineering in bridging generative models and downstream applications.

\subsubsection{The Impact of Input Modality}

To further clarify the impact of input modality on the effectiveness of LVLM-based representations, we conduct a model study to systematically compare different data modalities provided as input to the LVLM for micro-video recommendation. Specifically, we employed the hidden-state-centric feature extraction strategy on the MicroLens dataset, using three types of input modalities: video title, cover image, and raw video.

\begin{table}[t]
    \centering
    \caption{Performance comparison of different input modalities for VLLM-based micro-video recommendation on the MicroLens dataset.}
    \begin{tabular}{ccccc}
        \toprule
        Method & H@10 & N@10 & H@20 & N@20\\
        \midrule
        Title & 0.0773 & 0.0436 & 0.1076 & 0.0513\\
        Cover & 0.0862 & 0.0487 & 0.1197 & 0.0571\\
        Video & 0.0975 & 0.0545 & 0.1353 & 0.0640\\
        \bottomrule
    \end{tabular}
    \label{tab_input_modal}
\end{table}
\begin{table}[t]
    \caption{Case study comparing the effects of different input modalities on recommendation and ranking results.}
    \centerline{\includegraphics[width=\columnwidth]{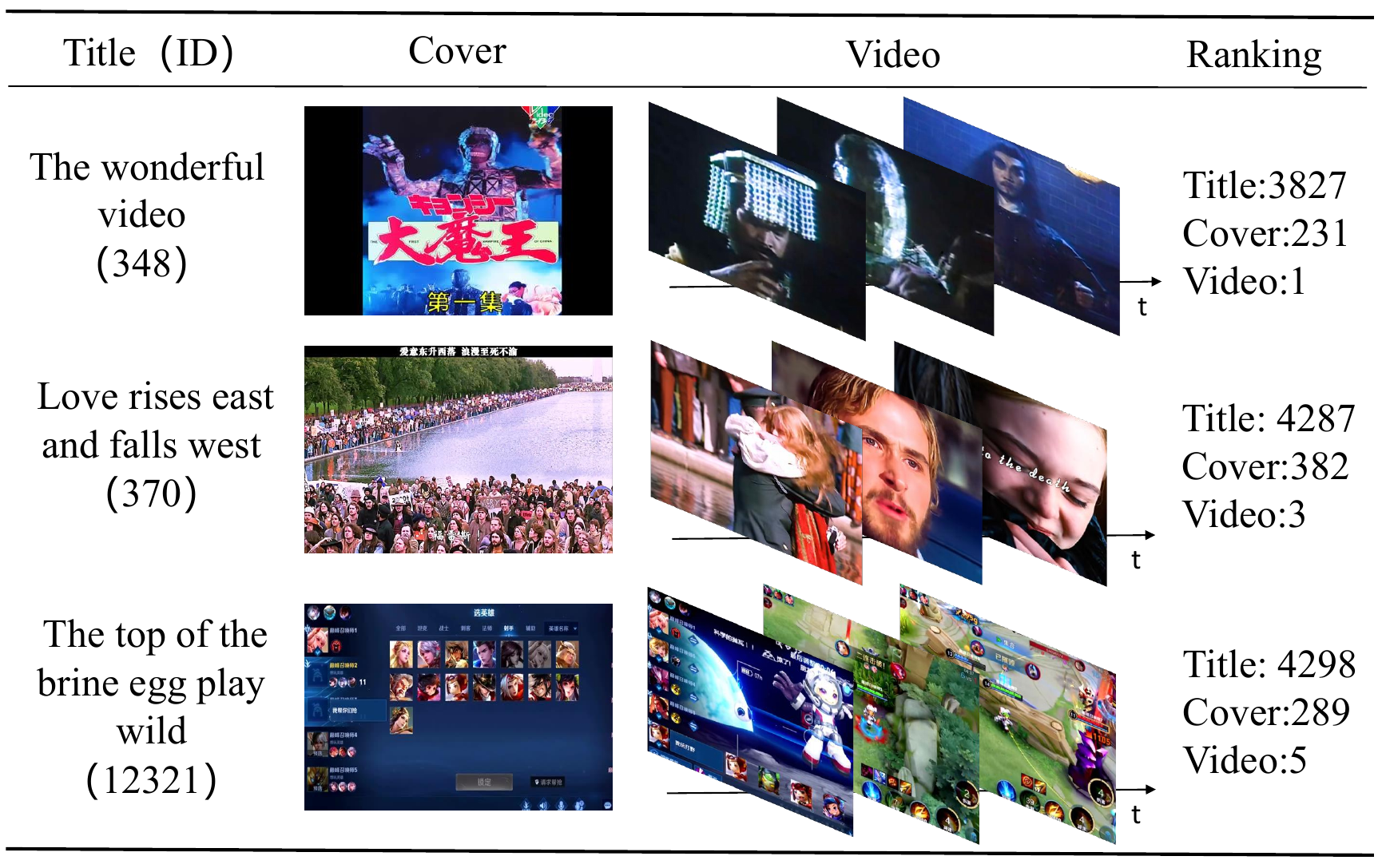}}
    \vspace{-10pt}
    \label{fig_casestudy}
\end{table}
As shown in Table~\ref{tab_input_modal}, representations from raw videos consistently outperform those from cover images and titles. This indicates that richer input modalities enable the LVLM to extract more discriminative content. This finding emphasizes the necessity of leveraging raw video content when constructing effective recommendation models based on LVLMs.

To more clearly illustrate the differences among the three input modalities, we conducted a case study by selecting three representative user sequences and presenting the corresponding target item’s data (title, cover image, and raw video) as well as the recommendation rankings. As shown in Figure~\ref{fig_casestudy}, micro-video titles are often insufficient or even misleading in capturing the actual video content, and covers cannot represent the full temporal dynamics and context. For instance, clickbait titles or static covers frequently lead to semantic drift, resulting in poor recommendation rankings. In contrast, leveraging raw video inputs enables the system to capture the true intent and contextual richness of micro-videos, allowing for more accurate target item identification. This confirms that raw videos, with their essential temporal and visual dynamics, are crucial for effective recommendation, especially when textual and static visual cues are ambiguous or incomplete.

\subsubsection{Training Efficiency and Convergence Analysis}
\begin{figure}[htbp]
    \centering
    \includegraphics[width=1\linewidth]{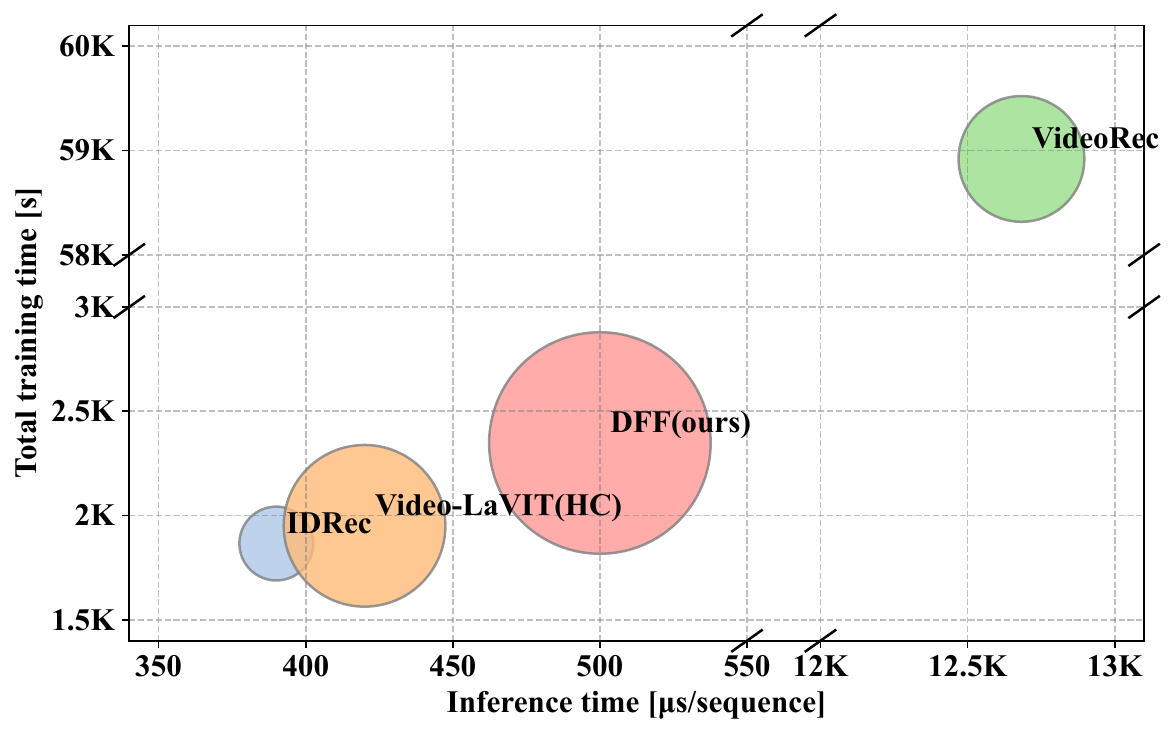}
    \caption{Comparison of total training time, inference time, and performance of models. The x-axis represents inference time per sequence (in \textmu s), the y-axis shows total training time (in seconds), and bubble radius indicates overall performance (larger is better). Video-LaVIT (HC) denotes the variant that extracts video features from the hidden states of the last decoder layer of Video-LaVIT.}
    \label{fig_efficiency}
\end{figure}

We record the total training time and inference time of DDF and the baseline methods, as reported in Figure~\ref{fig_efficiency}, where the bubble radius indicates each model’s overall performance. 
IDRec requires 1,866~s of training time and 390~\textmu s per inference sequence, while Video-LaVIT (HC) takes 1,951~s and 420~\textmu s, respectively. In comparison, DFF needs slightly more training time (2,348~s) and a modestly higher inference latency (500~\textmu s per sequence), yet it achieves the highest recommendation performance among all methods. In contrast, VideoRec has the longest training duration at 58,920~s and an excessively high inference time of 12,680~\textmu s, but still underperforms DFF. This demonstrates that DFF effectively leverages frozen large vision-language models to achieve an exceptional balance between effectiveness and efficiency.
Notably, DFF, IDRec, and Video-LaVIT (HC) are all trained on a single NVIDIA RTX 4090 GPU, whereas VideoRec, due to its substantial memory footprint and computational demands, is trained on two NVIDIA A800 GPUs. Despite this significantly more powerful hardware configuration, VideoRec not only converges much more slowly but also yields inferior recommendation performance, highlighting the remarkable efficiency and effectiveness of DFF.

\section{Conclusion and Future Work}

We present a systematic evaluation of key design choices for integrating frozen Large Video Language Models (LVLMs) into micro-video recommendation. Through extensive experiments on real-world datasets and representative LVLMs, we uncover three critical insights: 
(i) intermediate decoder hidden states consistently outperform caption-based representations, as they better preserve fine-grained visual semantics essential for recommendation; 
(ii) fusing LVLM features with ID embeddings significantly outperforms replacing ID embeddings, effectively leveraging the collaborative signals inherently captured by IDs; 
and (iii) adaptive fusion of multi-layer LVLM features yields further gains. 

Guided by these findings, we propose the Dual Feature Fusion (DFF) framework, a lightweight approach that adaptively combines multi-layer LVLM features with ID embeddings without requiring LVLM fine-tuning. DFF achieves state-of-the-art performance on the MicroLens and Bili\_Movie datasets, validating the effectiveness of our empirically grounded design principles. Our work shifts the paradigm from heuristic black-box usage of LVLMs toward transparent, insight-driven integration, providing actionable guidelines for deploying LVLM-based recommender systems.

Looking ahead, several promising directions merit further exploration. 
Although it is known that different Transformer layers capture varying levels of semantic knowledge and that multi-layer representations improve recommendation performance, how these features facilitate the learning of collaborative filtering signals is still not well understood. 
Moreover, while our framework adaptively weights multi-layer features, the question of which decoder layers to involve in the first place remains largely heuristic. Future work could systematically investigate principled layer selection strategies, such as choosing layers based on their semantic fidelity to user intent or interaction sparsity.

\begin{acks}
This research was supported by the Singapore Ministry of Education (MOE) Academic Research Fund (AcRF) Tier 1 grant.
\end{acks}

\bibliographystyle{ACM-Reference-Format}
\bibliography{refs}

\end{document}